\documentclass[12pt]{iopart}
\usepackage{graphicx}
\usepackage{subfigure}
\usepackage{multirow}
\usepackage[section]{placeins}
\usepackage{cases}
\usepackage{float} 
\usepackage[shortcuts]{extdash}
\usepackage{hyperref}
\usepackage{iopams}

\begin{document}

\title[Three-body bound states of two bosonic impurities]{Three-body bound states of two bosonic impurities immersed in a Fermi sea in 2D}

\author{F F Bellotti$^1$, T Frederico$^2$, M T Yamashita$^4$, D V Fedorov$^1$, A S Jensen$^1$ and N T Zinner$^1$}

\address{$^1$ Department of Physics and Astronomy, Aarhus University, DK-8000 Aarhus C, Denmark}
\address{$^2$ Instituto Tecnol\'{o}gico de Aeron\'autica, 12228-900, S\~ao Jos\'e dos Campos, SP, Brazil} 
\address{$^3$ Instituto de Fomento e Coordena\c{c}{\~a}o Industrial, 12228-901, S{\~a}o Jos{\'e} dos Campos, SP, Brazil}
\address{$^4$ Instituto de F\'isica Te\'orica, UNESP - Univ Estadual Paulista, 01156-970, S\~ao Paulo, SP, Brazil} 
\ead{bellotti@phys.au.dk}

\begin{abstract}
We consider two identical impurities immersed in a Fermi sea for a broad range of masses and for both interacting and non-interacting impurities. The interaction between the particles is described through attractive zero-range potentials and the problem is solved in momentum space. The two impurities can attach to a fermion from the sea and form three-body bound states.
The energy of these states increase as function of the Fermi momentum $k_F$, leading to three-body bound states below the Fermi energy. The fate of the states depends highly on two- and three-body thresholds and we find evidence of medium-induced Borromean-like states in 2D. The corrections due to particle-hole fluctuations in the Fermi sea are considered in the three-body calculations and we show that in spite of the fact that they strongly affect both the two- and three-body systems, the correction to the point at which the three-body states cease to exist is small.
\end{abstract}



\submitto{\NJP}

\maketitle

\section{Introduction}

The improvement of the techniques for cooling atoms has been boosting  advances in physics for several years~\cite{lewensteinAiP2007,blochRMP2008,esslingerARoCMP2010,zinnerJoPGNaPP2013}. A remarkable example is the finding of experimental evidences of the so-called Efimov states, that were predicted by V. Efimov in the seventies~\cite{efimovYF1970}. Using few-body techniques, he showed that the energy levels of three identical bosons accumulate with a geometrical separation between them when the binding energy of the pairs goes to zero. More than thirty years passed until a signature of such states was found in loss experiments in 2006~\cite{kraemerNP2006}. Although being a true breakthrough, this was still an indirect measurement and another nine years were necessary until this effect was found in a direct manner in experiments with $^{4}$He~\cite{kunitskiS2015}.

This impressive technological advance also opened new avenues as the quasi-2D realization of single-species systems~\cite{gorlitzPRL2001,burgerEL2002,hammesPRL2003}, two-component gases~\cite{martiyanovPRL2010,dykePRL2011}, mixed-species systems~\cite{modugnoPRA2003,gunterPRL2005} and even heteronuclear diatomic molecules~\cite{NPP2011} in ultracold atomic traps. On the theoretical side, it has been shown that three-identical weakly attractive bosons in 2D have only one two-body and two three-body bound states~\cite{bruchPRA1979} and further studies have also shown that the Efimov effect does not happen when the system is restricted to lower dimensions~\cite{limZfPAHaN1980,adhikariPRA1988}. Recent studies have focused on mixed-species two-dimensional systems~\cite{PricoupenkoPRA2010,bellottiJoPB2011,bellottiPRA2012}, which, despite of not supporting the Efimov states, can have a rich energy spectrum in the case of highly mass-imbalanced systems~\cite{bellottiJoPB2013,ngampruetikornEEL2013}.

While most of these calculations were done in vacuum or disregarding the medium, bound states can be drastically affected depending on the surrounding medium in which the system is immersed. If one thinks about a single impurity immersed in a bath of different atoms, its motion distorts the medium, which in turn acts back on the impurity. The final state depends on the interaction strength between the different species, where for weak enough attraction the impurity behaves as a free particle and for strong enough attraction it forms a bound state with a particle from the sea. On the other hand, for intermediate attraction the impurity is dressed by the particles from the sea and becomes a polaron, namely, a quasi-particle whose basic properties as effective mass and charge differs from the ones of the original impurity. These different properties result from the dressing of the impurity by excitations of the background medium. The original idea of polaron dates back to Landau~\cite{landauPZS1933} and a discussion about it in the context of ultracold atomic gases can be found in~\cite{massignanRoPiP2014}.

It is possible to simulate polaronic systems with cold atomic gases by mixing a few atoms of a specific type to a bath of atoms of a different kind~\cite{chikkaturPRL2000} (different masses or spin states, for example) even when the system is restricted to two dimensions and the majority atoms are fermions, in which case the quasi-particle is known as a Fermi polaron. Fermi gases have been successfully trapped to quasi-2D geometries and the Fermi polaron was already observed through Radio-Frequency Spectroscopy techniques~\cite{schirotzekPRL2009,frohlichPRL2011,zhangPRL2012,kohstall2012}. The fate of a single impurity immersed in a two-dimensional Fermi sea has been studied in recent years both experimentally~\cite{koschorreck2012} and theoretically~\cite{zollnerPRA2011,parishPRA2011,schmidtPRA2012,ngampruetikornPRL2013,parishPRA2013}, where the importance of the particle-hole fluctuations in 2D systems was emphasized. 

In real systems it can be very hard to have only a single particle propagating in the medium and it is therefore important to understand how the presence of extra impurities affect the system. In this work we take a first step in this direction and consider the case of two identical impurities immersed in a Fermi sea. The study applies to a broad range of masses and both for interacting and non-interacting impurities. In both cases, the two impurities can attach to a fermion from the sea and form three-body bound states. We show how the energy of the states are shifted depending on the Fermi momentum $k_F$ and also how these states decay into the two- and three-body continuum. The corrections due to the fluctuations in the sea are considered when the impurities are not allowed to interact with each other. We show that in spite of the strong influence both the two- and three-body systems, the corrections from fluctuations in the Fermi sea have small influence on the point where the three-body bound state cease to exist. 

This paper is organized as follows: the influence of the background environment in the state of an impurity bound to a fermion from the sea is discussed in Sec.~\ref{sec:bvstbs} and the derivation of the equations which consider corrections to the two-body system due to the particle-hole fluctuations in the sea are presented in Sec.~\ref{sec:sectbsp}. Sec.~\ref{sec:ttbs} brings a discussion about the dependence of the two- and three-body thresholds in the Fermi momentum. Three-body systems composed of two identical interacting impurities and a fermion from the sea are studied in Sec.~\ref{sec:ii}. Corrections to the three-body states induced by the fluctuations in the sea are discussed in Sec.~\ref{sec:niisectbs} for two non-interacting particles, where it is also argued how the corrections would affect the three-body system when the impurities do interact. Discussion and outlook are in Sec.~\ref{sec:do} and technical details are given in \ref{sec:ese} and \ref{sec:apie}.

\section{Bound and virtual states of the two-body system} \label{sec:bvstbs}

We consider an atom labeled $a$ of mass $m_a$ interacting with a fermion $b$ of mass $m_b$ that is on top of a background Fermi sea of momentum $\vec{k}_F$. 
The interaction is described for the one term separable potential $V=\lambda \left|\chi\right\rangle \left\langle\chi\right|$, it is attractive ($\lambda<0$) and has the form factor $g(p)=\left\langle \vec{p} \right| \left. \chi\right\rangle=1$  because we use a Dirac delta potential.
The matrix elements of the two-body T-matrix~\cite{adhikariAJoP1986,bellottiJoPB2011} under the influence of the Pauli blocking effects are calculated as 
\begin{equation}
\tau^{-1}(E_2)=\lambda^{-1} - \int{d^2p \frac{\Theta \left( \left| \vec{p}-\frac{m_{ab}}{m_a} \vec{q} \right| - k_F \right)}{E_{2}-\frac{p^2}{2 m_{ab}}+i  \epsilon} } , 
\label{eq01}
\end{equation}
where $\vec{p}$ is the relative momentum, $\vec{q}$ is the center-of-mass (CM) momentum of the pair with respect to the Fermi sea, the reduced mass is $m_{ab}=m_a m_b/(m_a+m_b)$ and $E_2$ is the internal (or binding) energy of the pair. The $\Theta-$function ensures that the fermion is not inside the Fermi sea. The divergence in (\ref{eq01}) is treated with the subtraction method~\cite{adhikariPRL1995}, where the strength of the potential $\lambda$ is connected to the binding energy of the pair $|E_{ab}|$ through $\lambda^{-1}=-\int d^2p \left(|E_{ab}|+p^2/2 m_{ab}\right)^{-1}$. Performing the integral in (\ref{eq01}) yields
\begin{eqnarray}
\fl \tau^{-1}(E_2)=-2 \pi m_{ab} \ln\left( \frac{\left[\frac{1}{2 m_{ab}} \left(k_F-\frac{m_{ab}}{m_a} q\right)^2-E_{2}\right]^{1/2} \left[\frac{1}{2 m_{ab}} \left(k_F+\frac{m_{ab}}{m_a} q\right)^2-E_{2}\right]^{1/2}}{2 |E_{ab}|} \right. \nonumber\\* \left. +\frac{ \frac{k_F^2}{2 m_{ab}}  -\frac{m_{ab}}{2 m_a^2} q^2 -E_{2} }{2 |E_{ab}|}\right)  . 
\label{eq06}
\end{eqnarray}
Notice that if the background Fermi sea is absent, $k_F=0$ and (\ref{eq06}) recovers the two-body T-matrix elements of an $ab$ system in vacuum, namely
\begin{equation}
\tau^{-1}(E_2)=-4 \pi m_{ab} \ln\left( \sqrt{\frac{-E_2}{|E_{ab}|}} \right)   \; .
\label{eq12}
\end{equation}

It is possible to identify three different thresholds in (\ref{eq06}). As discussed in~\cite{schmidtPRA2012} for $m_a=m_b$, two of them come from the branch cut of the square root and are given by $E_{th}^\pm=\frac{1}{2 m_{ab}} \left(k_F \pm \frac{m_{ab}}{m_a} q\right)^2$. The third one arises from the branch cut of the logarithm and reads $E_{th}^0=0$ for $q \geq \frac{m_a}{m_{ab}} k_F$.

The two-body system only supports bound states when $(i)$ there is a pole in the two-body T-matrix and $(ii)$ the energy where the pole appears is below the lowest threshold, namely $E_2<E_{th}^-$ or $E_2<E_{th}^0$ for $q \geq \frac{m_a}{m_{ab}} k_F$. Setting (\ref{eq06}) equal to zero, the two-body energies corresponding to bound states are described by the expression
\begin{equation}
E_2=-|E_{ab}|+\frac{k_F^2}{2 m_{ab}}\frac{1}{1+\frac{m_{ab}}{2 m_a^2}\frac{q^2}{|E_{ab}|}},
\label{eq10}
\end{equation}
which tends to the energy of the pair in vacuum when the Fermi sea is not relevant, i.e., $E_2 \to -|E_{ab}|$ for $q/k_F \to \infty$. This expression agrees with the ones calculated using many-body techniques, variational methods etc.~\cite {zollnerPRA2011, parishPRA2011, schmidtPRA2012,ngampruetikornEEL2012,parishPRA2013}. 

\begin{figure}[!htb]%
\subfigure[\ $k_F < \sqrt{8 m_{ab} |E_{ab}|}$ \label{energy1}]{
\includegraphics[width=0.49\columnwidth]{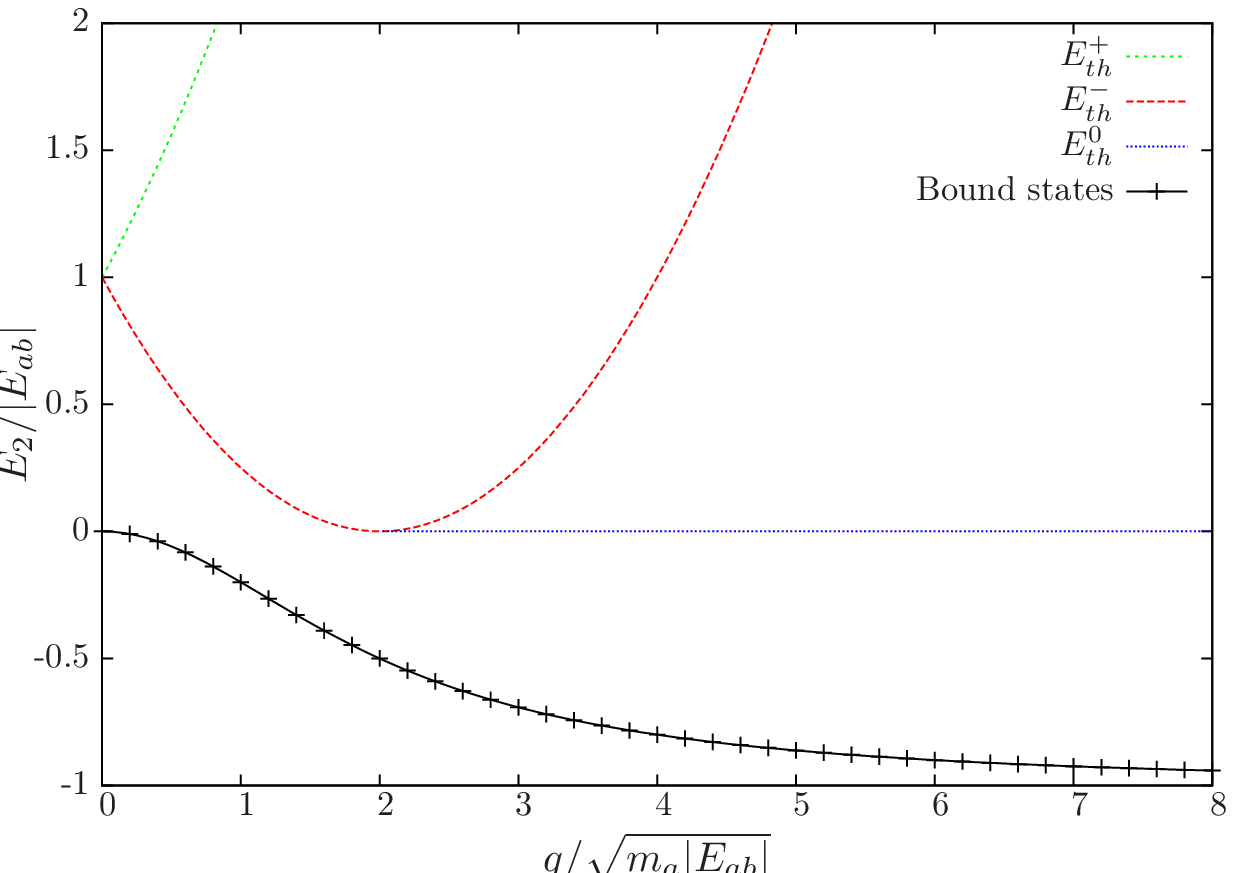}}
\subfigure[\ $k_F > \sqrt{8 m_{ab} |E_{ab}|}$ \label{bound-virtual}]{
\includegraphics[width=0.49\columnwidth]{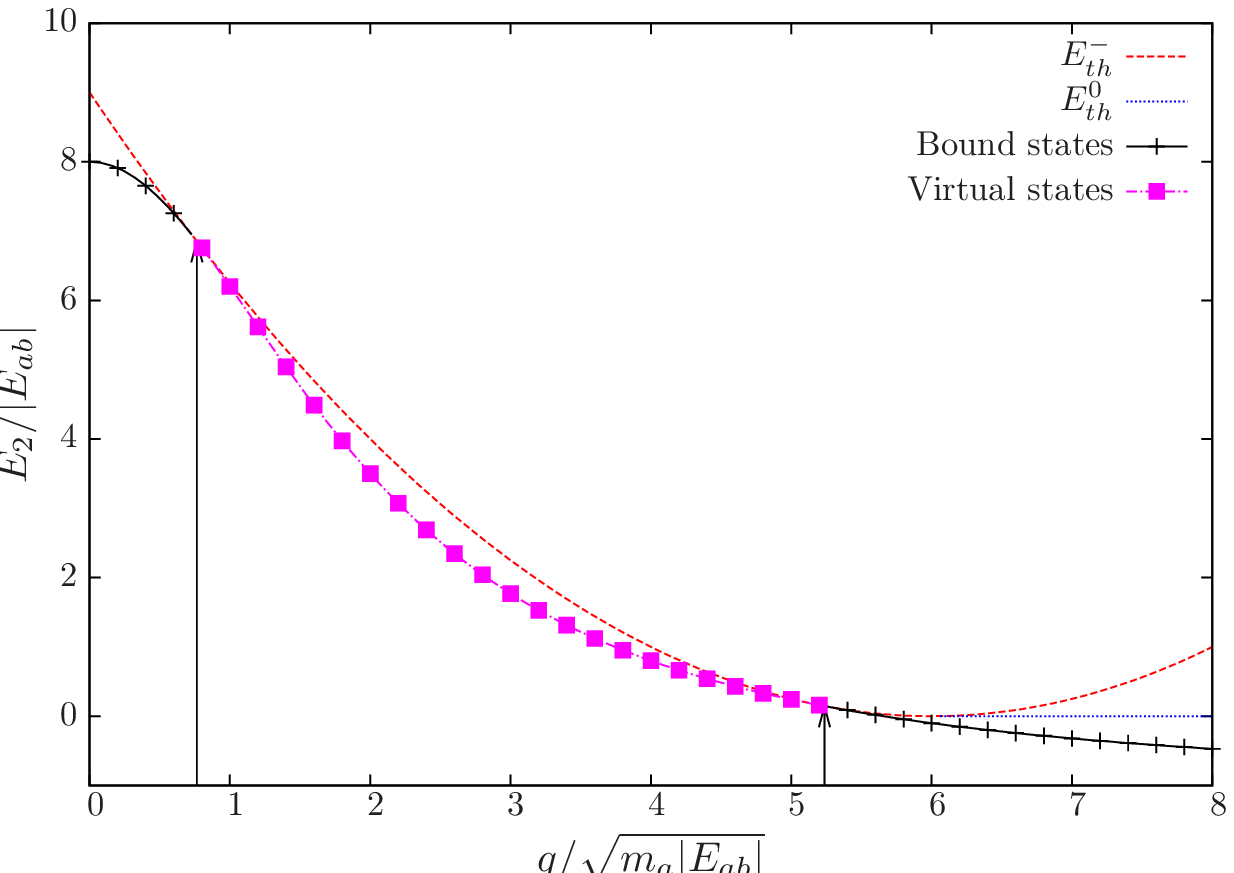}}

\caption{Two-body internal energy, $E_2/|E_{ab}|$, as function of the CM momentum $q/\sqrt{m_a |E_{ab}|}$. The analytic expression for the two-body energy in (\ref{eq10}) perfectly matches numerical calculation for both bound and virtual states. \ref{energy1}: Only bound states are present for $k_F < \sqrt{8 m_{ab} |E_{ab}|}$. \ref{bound-virtual}: Both bound and virtual states are present for $k_F > \sqrt{8 m_{ab} |E_{ab}|}$. The vertical arrows are the extremes of (\ref{eq11}).}
\label{bound-virtual-continuation}%
\end{figure}

The internal two-body energy given in (\ref{eq10}) always satisfies the condition $(ii)$, but surprisingly it does not satisfy the condition $(i)$ for all possible combinations of momenta $\vec{q}$ and $\vec{k}_F$. There is a strong competition between binding, the Fermi sea and the CM momentum such that the pair is unbound for $k_F > \sqrt{8 m_{ab} |E_{ab}|}$ when the value of these parameters fulfill the relation
\begin{equation}
\frac{m_a}{2 m_{ab}} \left(k_F-\sqrt{k_F^2-8 m_{ab} |E_{ab}|}\right) \leq q \leq  \frac{m_a}{2 m_{ab}} \left(k_F+\sqrt{k_F^2-8 m_{ab} |E_{ab}|}\right)   .
\label{eq11}
\end{equation} 
Although the analytical expression (\ref{eq10}) is a solution of $\tau^{-1}(E_2)=0$, numerical calculations from (\ref{eq01}) also do not indicate two-body bound states in the region given in (\ref{eq11}), which has its boundaries framed by the vertical arrows in figure~\ref{bound-virtual}. 

The absence of bound states in the region defined by (\ref{eq11}) was already pointed out in~\cite{schmidtPRA2012} and can also be seen in~\cite{EngelbrechtPRB1992} for the spin-balanced system and the remaining question is to know what happens to the state in this region. It is shown in figure~\ref{bound-virtual-continuation} that for $k_F > \sqrt{8 m_{ab} |E_{ab}|}$ the state touches the lowest threshold ($E_{th}^{-}$) in two points, strongly suggesting that it enters through the two-body cut in one point and comes back in another one. This cut is determined exclusively by the square-root inside the logarithm in (\ref{eq06}) and the logarithm itself does not contribute to the cut since its argument is positive and non-zero throughout the regions where $E_2 < E_{th}^-$ and $E_2<E_{th}^0$ for $q \geq \frac{m_a}{m_{ab}} k_F$. 

The analytic continuation of the matrix elements (\ref{eq06}) is labeled $\tau^{-1}_*$ and found by flipping the sign in front of the square-root inside the logarithm in (\ref{eq06}), since the logarithm itself does not contribute to the cut. Next, this expression is used to calculate $E_2$ as a solution of  $\tau^{-1}_*(E_2)=0$, which gives precisely (\ref{eq10}) as the result. Notice that now the solution still satisfies condition $(ii)$, but only satisfies condition $(i)$ in the region given in (\ref{eq11}) and not outside this region as before. These two-body states have entered through the cut and are in the second energy-sheet, therefore they are virtual states. These states are shown in figure~\ref{bound-virtual}. 


The matrix elements of the two-body T-matrix as given in (\ref{eq06}) may lead to inconsistent results in the scattering region ($E_2>E_{th}^{-}$), where the imaginary part of the energy becomes important. For instance, setting $E_2 \to E_2+i  \epsilon$ in (\ref{eq06}), expanding the terms inside the square root  and collecting energies and $\epsilon$'s together makes it possible to see that the imaginary part becomes null in some cases. 
If one disregards for a moment the small imaginary term that should appear inside the square roots and considers that the imaginary contribution due to these terms come only from the arguments inside it, then three cases have to be considered, namely, $E_2 < E_{th}^-$, $E_{th}^- \leq E_2 \leq E_{th}^+$ and $E_2 \geq E_{th}^+$. 

When the energy crosses from $E_2 < E_{th}^-$ to $E_2 \geq E_{th}^-$, the analytic extension of the square root has to be properly taken in account. In other words, its imaginary part has to follow the sign of the $i  \epsilon$ term that appears outside the square roots in (\ref{eq06}), meaning that $\left[1/2 m_{ab} \left(k_F-m_{ab}/m_a \ q\right)^2-E_{2}\right]^{1/2} \to -i  \left[E_{2}-1/2 m_{ab} \left(k_F-m_{ab}/m_a \ q\right)^2\right]^{1/2}$ for $E_2>E_{th}^{-}$. 
The same argument is taken in the transition from $E_{th}^- \leq E_2 \leq E_{th}^+$ to $E_2 \geq E_{th}^+$ and therefore, the matrix elements are written as
\begin{equation}
\tau^{-1}(E_2)=-2 \pi m_{ab} \ln\left( \frac{g(E_2,k_F,q) + \frac{1}{2 m_{ab}} \left( k_F^2 - \frac{m_{ab}^2}{m_a^2} q^2 \right)-E_{2}-i  \epsilon }{2 |E_{ab}|} \right)  , 
\label{eq08}
\end{equation}
with $g(E_2,k_F,q)$ given by
\begin{equation}
\left[\frac{1}{2 m_{ab}} \left(k_F-\frac{m_{ab}}{m_a} q\right)^2-E_{2}\right]^{1/2} \left[\frac{1}{2 m_{ab}} \left(k_F+\frac{m_{ab}}{m_a} q\right)^2-E_{2}\right]^{1/2}    \label{eq09a1}
\end{equation}
for $E_2 < E_{th}^-$,
\begin{equation}
-i  \left[E_{2}-\frac{1}{2 m_{ab}} \left(k_F-\frac{m_{ab}}{m_a} q\right)^2\right]^{1/2} \left[\frac{1}{2 m_{ab}} \left(k_F+\frac{m_{ab}}{m_a} q\right)^2-E_{2}\right]^{1/2}   \label{eq09b1} 
\end{equation}
for $E_{th}^- < E_2 < E_{th}^+$ and finally
\begin{equation}
- \left[E_{2}-\frac{1}{2 m_{ab}} \left(k_F-\frac{m_{ab}}{m_a} q\right)^2\right]^{1/2} \left[E_{2}-\frac{1}{2 m_{ab}} \left(k_F+\frac{m_{ab}}{m_a} q\right)^2 \right]^{1/2}  \label{eq09c1}
\end{equation}
for $E_2 > E_{th}^+$.

\begin{figure}[!htb]%
\centering
\includegraphics[width=0.8\columnwidth]{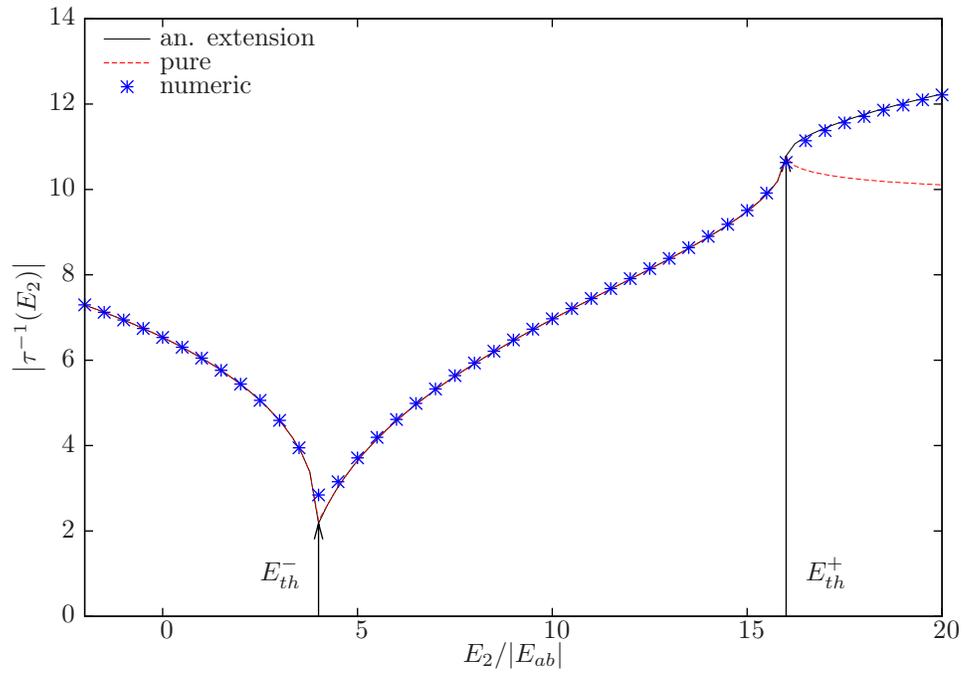}
\caption{Comparison between the absolute value of the matrix elements with the analytic extension in (\ref{eq08}), with $E_2 \to E_2+i  \epsilon$ in (\ref{eq06}) and numerically calculated from (\ref{eq01}) as function of $E_2/|E_{ab}|$ for $m_a=m_b$, $q/\sqrt{m_a |E_{ab}|}=2$ and $k_F/\sqrt{m_a |E_{ab}|}=3$ .}%
\label{tau}%
\end{figure}

A comparison between the absolute value of the matrix elements with the analytic extension in (\ref{eq08}), with  $E_2 \to E_2+i  \epsilon$ in (\ref{eq06}) and numerically calculated from (\ref{eq01}) is shown in figure~\ref{tau}. The analytic extension becomes more relevant as $E_2$ approaches $E_{th}^+$. For $E_2<E_{th}^+$ the analytic form in (\ref{eq06}) nicely reproduces the numerical calculation.

\begin{figure}[!htb]%
\subfigure[\ \label{fig09a} $\frac{E_{ab}}{E_F}=0.1$, without analytic extension.]
{\includegraphics[width=0.5\columnwidth]{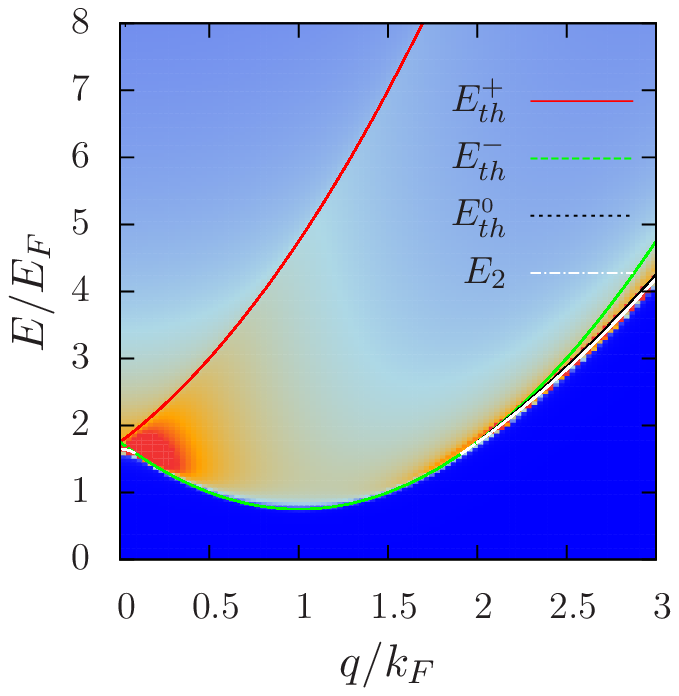}}
\subfigure[\ \label{fig09b} $\frac{E_{ab}}{E_F}=0.1$, with analytic extension.]
{\includegraphics[width=0.5\columnwidth]{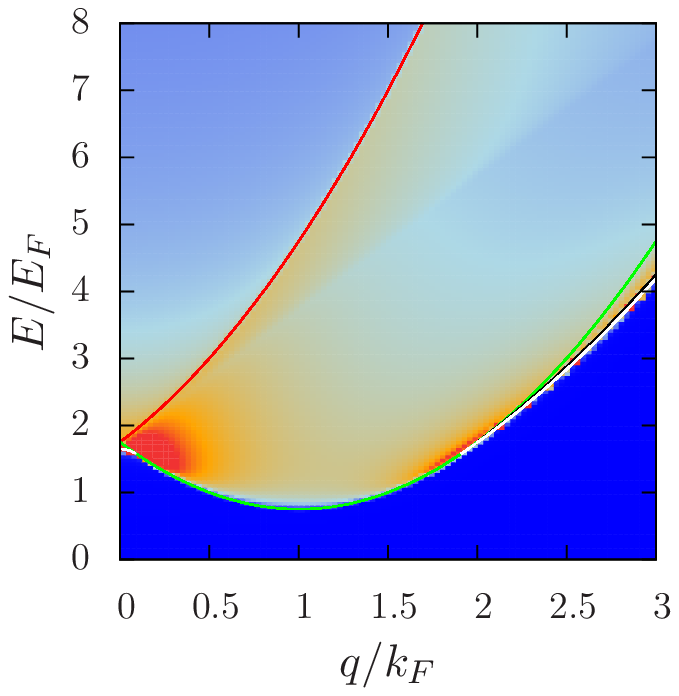}}

\subfigure[\ \label{fig09c} $\frac{E_{ab}}{E_F}=2.5$, without analytic extension.]
{\includegraphics[width=0.5\columnwidth]{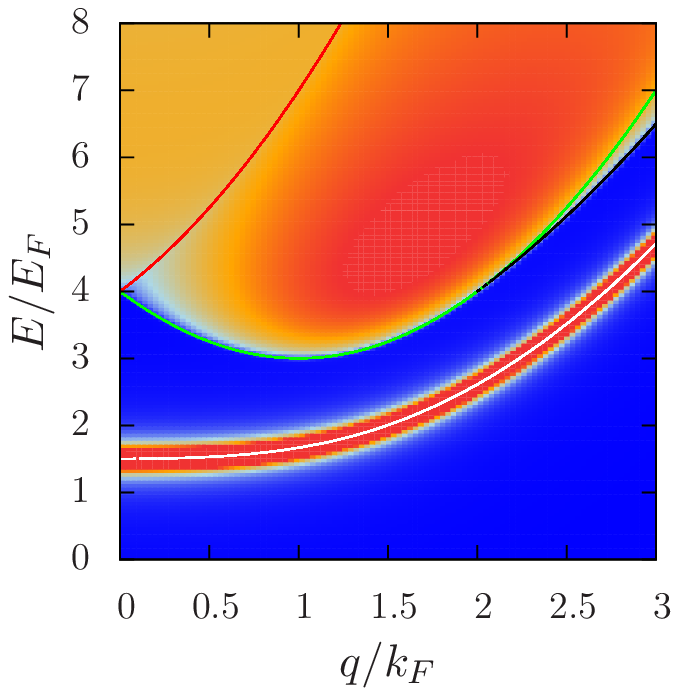}}
\subfigure[\ \label{fig09d} $\frac{E_{ab}}{E_F}=2.5$, with analytic extension.]
{\includegraphics[width=0.5\columnwidth]{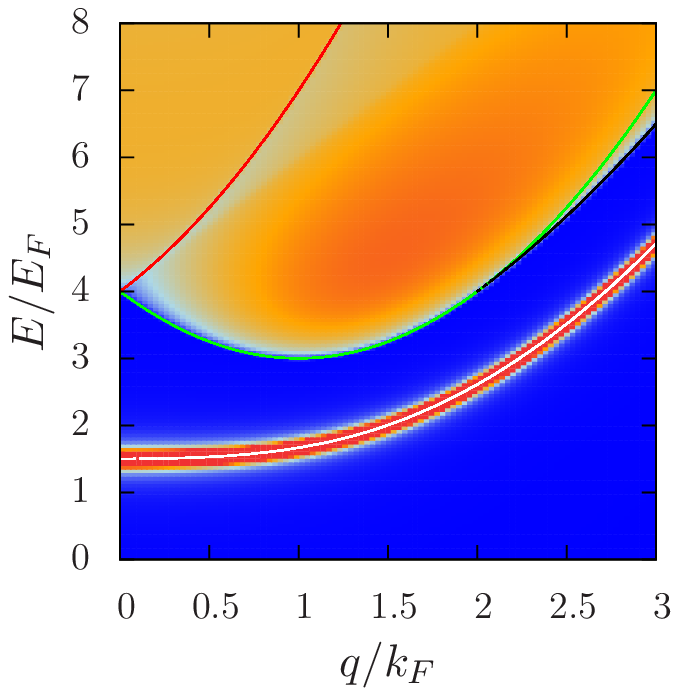}}
\caption{Spectral molecular function $A_{mol}(E,q)$ for $\frac{E_{ab}}{E_F}=0.1$ (top) and $\frac{E_{ab}}{E_F}=2.5$ (bottom). The left-hand-side is the result of calculations with \eref{eq06} and is the same as presented in~\cite{schmidtPRA2012}. The right-hand-side is calculated with \eref{eq08}, where the analytic extension of the matrix elements has been properly taken into account. Notice how the contrast change from left to right and that the corrected expression for the matrix elements is more relevant in the large binding limit.    }%
\label{spectral-mol}%
\end{figure} 

The correction introduced by the analytic extension can be seen in the spectral molecular function, defined as $A_{mol}(E,q)=-2 \Im \;\tau(E,q)$. Figures~\ref{fig09a} and \ref{fig09c} show the spectral molecular function respectively for $\frac{E_{ab}}{E_F}=0.1$ and $2.5$ calculated with \eref{eq06}, where a small imaginary part is added to the two-body energy. On the other hand, figures~\ref{fig09b} and \ref{fig09d} are calculated with \eref{eq08}, where the analytic extension of the matrix elements has been properly taken into account. Notice that the use of the proper expression slightly changes the contrast in the figures. Figures~\ref{fig09a} and \ref{fig09c} have been obtained in~\cite{schmidtPRA2012} and in order to straightforwardly compare figure~\ref{spectral-mol} to the previous work, calculations here were made replacing the binding energy in \eref{eq06} and \eref{eq08} by the total two-body energy measured in terms of the Fermi energy $E_F$ plus the chemical potential of the impurity ($\mu_a$), specifically $E_2 \to E_2-\frac{q^2}{2(m_a+m_b)}+\mu_a+E_F$ on those equations.

\section{Self-energy correction to the two-body system and Polaron} \label{sec:sectbsp}
A way of taking into account fluctuations in the background medium is by introducing the self-energy of the impurity, which considers particle-hole excitations on top of the Fermi sea. The full propagator for the self-energy of the impurity relates to the bare one through the Dyson equation 
\begin{equation}
G_\Sigma^{-1}(E,\vec{p})=E-\frac{p^2}{2m_a}-\Sigma_a(E,\vec{p}).
\end{equation}

In the self-consistent way, the transition operator and the self-energy obey a set of non-linear coupled equations. This problem is handled by solving this set of non-linear coupled equations iteratively, which is equivalent to a Ward-Luttinger approach~\cite{LuttingerPR1960}. The zero-order transition operator is the bare one, given in (\ref{eq08}) and the zero-order self-energy is a function of the bare transition operator. The first-order transition operator is a function of the zero-order self-energy and so on (see \ref{sec:ese} for the details). This procedure leads to the set of coupled equations  
\begin{eqnarray}
\fl \Sigma_a^n(E_a,\vec{p}_a)=\int_{p_b<k_F}{d^2p_b \; \tau_n\left(\left|\vec{p}_b+\vec{p}_a\right|,\frac{p_b^2}{2 m_b}+E_a\right)} \; , \; n \geq 0 \label{ts1} \\
\fl \tau^{-1}_{n+1}(E_2,q)=\lambda^{-1} \nonumber\\*
\hskip -5em -\int{ d^2k 
\frac{\Theta \left( k - k_F \right)}{-E_2 +\frac{k^2}{2 m_{ab} }+\frac{m_{ab}}{2 m_a^2} q^2+ \frac{k q}{ m_a} \cos\theta + \Sigma_a^n(E_2+\frac{q^2}{2(m_a+m_b)}-\frac{k^2}{2 m_b},\vec{q}-\vec{k}) - i  \epsilon}} \;  , \label{ts}
\end{eqnarray} 
where $\tau^{-1}_{0}(E_2,q)$ is given in (\ref{eq08}), $m_a$ is the mas of the impurity and $m_b$ the mass of the fermion.

The problem of one single impurity on top of a Fermi sea has been successfully solved with non-self-consistent methods (see, e.g.~\cite{schmidtPRA2012,parishPRA2013,ngampruetikornEEL2012,vlietinckPRB2014,LevinsenPRA2012}). However, we found the self-consistent one more suitable to handle in the investigation of the three-body problem of two impurities immersed in a Fermi sea, whose results are presented in the following sections. This choice is supported since the self-consistent method describes well the two-body system when particle-hole fluctuations are taken into account. For instance, calculations done with the zero-order expression of the self-energy (\ref{ts1} with $n=0$) for $m_a=m_b$ reproduce previous results in the literature. Using (\ref{ts1}) and (\ref{ts}) we found that the energy of the attractive and repulsive branches of the polaron, which are solutions from $E=q^2/\left(2 m_a\right)+\Re\left[ \Sigma_a^0(E,\vec{q})\right]$, as well as the weights of the branches  ($Z=\left\{1-\partial_{E} \left[ \Re \left(\Sigma_a^0(E,\vec{q}) \right) \right] \right\}^{-1}$ ) and the polaronic spectral function ($A_a(E,\vec{q})=-2 \Im[E+i  \epsilon - \Sigma_a^0(E,\vec{q})]^{-1} $) are the same as found in~\cite{ngampruetikornEEL2012,schmidtPRA2012}.

In~\cite{parishPRA2013,vlietinckPRB2014} the molecular regime, namely the region where the energy of the molecule under the effect of particle-hole fluctuations from the medium (dressed molecule) is lower than the polaron one, calculated with non-self-consistent methods happens from large and negative $\eta$ up respectively to -0.97 or -0.95,  above which the polaron energy will always be below the molecular energy. We define $\eta=\ln\left(k_F a_{2D}\right)=-\frac{1}{2} \ln\left(\frac{|E_{ab}|}{E_F} \frac{m_{ab}}{m_b}\right)$, $a_{2D}$ is the two-dimensional scattering length which fulfills $E_{ab}=\left(2 m_{ab} a_{2D}^{2}\right)^{-1}$ and $E_F=k_F^2/2 m_b$ is the Fermi energy.  This transition between regimes is often called polaron-molecule transition and we found a first-order ($n=0$ in (\ref{ts1}) and (\ref{ts}) ) polaron-molecule transition to occur around $\eta=-0.54$, which is below the results obtained with variational~\cite{parishPRA2013} and diagrammatic Monte Carlo~\cite{vlietinckPRB2014} calculations, but is close to the $\eta \approx -0.60$ found in the 2D limit of quasi-two-dimensional highly polarized Fermi gases~\cite{LevinsenPRA2012}.

The different behavior found between our method and previous studies in the small window  $-0.96(\pm 0.01)< \eta <-0.54$~\cite{parishPRA2013,vlietinckPRB2014} or $-0.60< \eta <-0.54$~\cite{LevinsenPRA2012} is not important for our purpose of using the self-consistent approach in the three-body calculations. The actual point of crossing is not the relevant information since experiments~\cite{koschorreck2012} have shown that the crossing of the energies of the polaron and the molecule as function of $\eta$ happens at a very shallow angle and that the two-body energy spectrum is not affected by the polaron-molecule transition (see~\cite{koschorreck2012} for details).

\begin{figure}[!htb]%
\centering
\includegraphics[width=0.8\columnwidth]{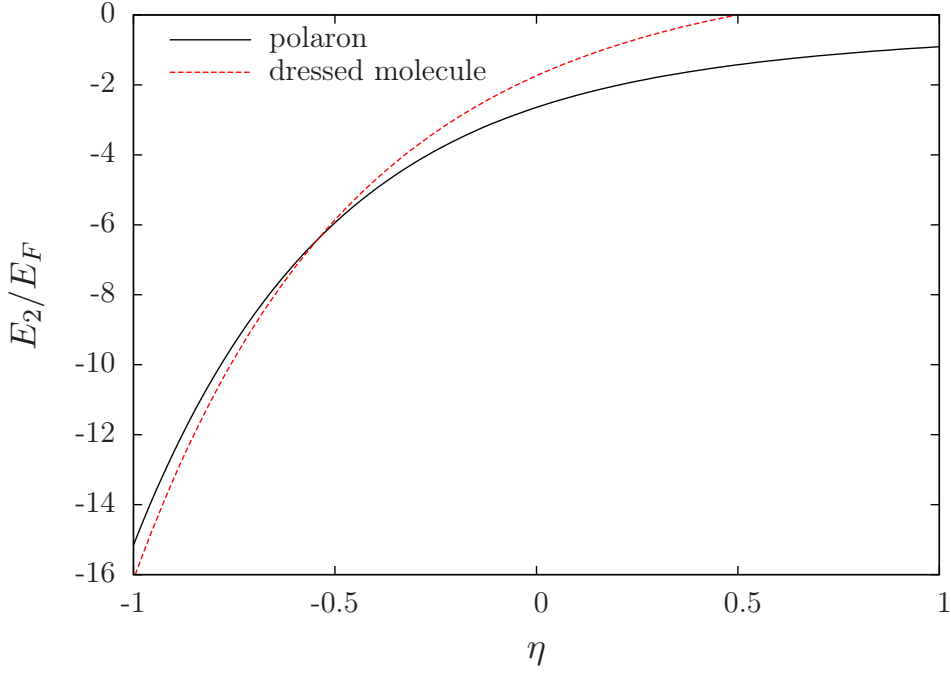}
\caption{The energy spectrum of the polaron (attractive branch) and the dressed molecule as function of the parameter $\eta$. Notice that the crossing between lines happens at a shallow angle and that both spectra are similar in the region $-1.0<\eta<-0.54$, from where the polaron has always a lower energy. }%
\label{fig10}%
\end{figure}

In view of that, it is more important that the model reproduces this transition than the actual point of occurrence. Solving (\ref{ts1}) and (\ref{ts}) with $n=0$ we find that the energy spectrum of the polaron and the dressed molecule  are very similar to each other and that the crossing between the energies does happen at a shallow angle, as shown in figure~\ref{fig10}.  Here a difference between the self-consistent method and the non-self-consistent one can be seem. Comparing figure~\ref{fig10} to~\cite{LevinsenPRA2012}, we note that the polaronic spectrum is exactly the same, while the spectrum of the dressed molecule has a small quantitative difference. However, as explained above, since our model captures the correct behavior of the two-body system, this small quantitative difference is not a problem. As an extra argument for using the self-consistent method in our analysis, notice that our three-body results when particle-hole fluctuations are taken into account lie mostly in $\eta>-0.54$ (see figures~\ref{diagrams} and \ref{diagram_corrected}), from where the energy of the polaron is always lower than the molecular one and it is the same as calculated with non-self-consistent methods~\cite{schmidtPRA2012,parishPRA2013,ngampruetikornEEL2012,vlietinckPRB2014,LevinsenPRA2012}.

It was found in~\cite{vlietinckPRB2014}, with diagrammatic Monte Carlo calculation, that corrections beyond one particle-hole in the medium does not contribute significantly to the energy spectrum of the polaron and the dressed molecule. In view of that, the convergence of the iterative solution of (\ref{ts1}) and (\ref{ts}) is studied. $\Sigma_a^n(E_a,0)$ and $\tau^{-1}_{n+1}(E_2,0)$ are shown as function of $E_2/E_{ab}$ for different $\eta$'s, where the strong binding-limit corresponds large negative $\eta$ and the weak binding-limit to large positive $\eta$.  

\begin{figure}[!htb]%
\subfigure[\ $\eta=-2$ \label{tausigma_q0_eta-2} \label{sigmatau-2}]{
\includegraphics[width=0.49\columnwidth]{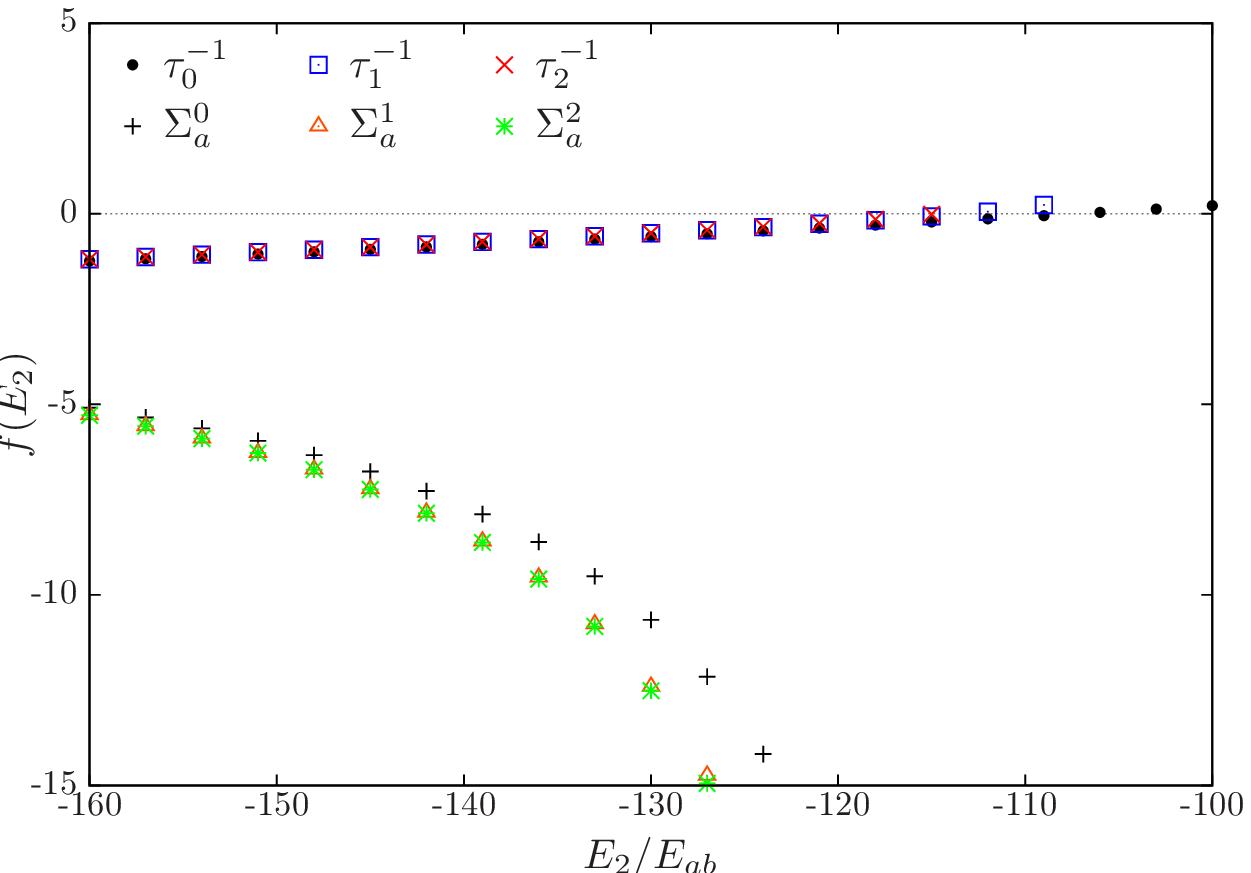}}
\subfigure[\ $\eta=0$ \label{tausigma_q0_eta+0} \label{sigmatau+0}]{
\includegraphics[width=0.49\columnwidth]{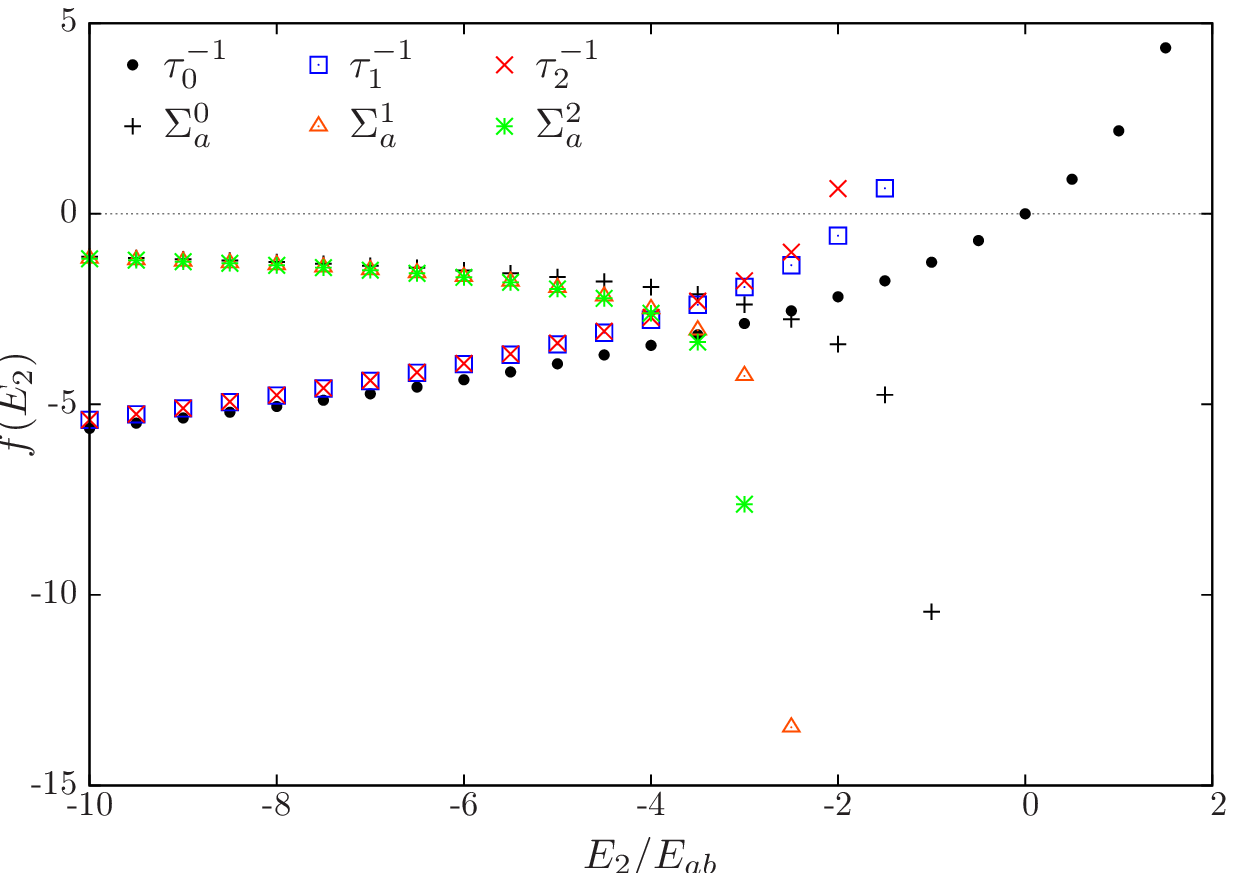}}
\caption{The transition operator $\tau_n^{-1}$ and the self-energy $\Sigma_a^n$ as functions of $E_2/E_{ab}$ for $n=0,1,2$ and  $q/k_F=0$.}
\label{sigmatau}
\end{figure}

Results for large binding ($\eta=-2$) are shown in figure~\ref{sigmatau-2}. Notice that the self-energy converges very fast as $\Sigma^1_a(E_2)$ and $\Sigma^2_a(E_2)$ are almost indistinguishable from each other and both of them differ just slightly from $\Sigma^0_a(E_2)$. As expected, the self-energy correction is almost negligible in this limit, which is confirmed by noticing that $\tau_0^{-1}(E_2)$, $\tau_1^{-1}(E_2)$ and $\tau_2^{-1}(E_2)$ are practically identical.

Increasing $k_F$ and moving towards the small binding limit, a few iterations are still needed for the self-energy to converge. For $\eta=0$, for instance, the difference from $\Sigma^2_a(E_2)$ to $\Sigma^1_a(E_2)$ is clearly much smaller than the difference from $\Sigma^1_a(E_2)$ to $\Sigma^0_a(E_2)$, as shown in figure~\ref{sigmatau+0}. The difference is that now the self-energy influences the two-body system, rendering it more bound. There is a big difference between the pole position of $\tau_0^{-1}(E_2)$ and $\tau_1^{-1}(E_2)$, but this difference is already much smaller from $\tau_1^{-1}(E_2)$ to $\tau_2^{-1}(E_2)$.

The scenario keeps this pattern for $\eta>0$ and figures showing details are not presented here. 
The results in figure~\ref{sigmatau} provide our argument for working with $\tau_1^{-1}(E_2)$ and $\Sigma^0_a(E_2)$ in the three-body calculations below. The computational cost of going beyond this are considerable and we do not expect neither qualitative nor large quantitative changes.

One great simplification comes by noticing that the function $\Sigma$ is almost independent of the angles in its argument, namely
\begin{eqnarray}
\hskip -5em \int_0^{2 \pi}{ d\theta
\frac{1}{f_1(E_2,q,k) + \frac{k q}{ m_a} \cos\theta + \Sigma_a^n(f_2(E_2,q,k),\vec{q}-\vec{k})}} \approx \nonumber\\*  
\hskip +2em \int_0^{2 \pi}{ d\theta
\frac{1}{f_1(E_2,q,k) + \frac{k q}{ m_a} \cos\theta + \Sigma_a^n(f_2(E_2,q,k),q^2+k^2)}} \ ,
\label{sigma-app2}
\end{eqnarray}
where $f_1(E_2,q,k)=-E_2 +\frac{k^2}{2 m_{ab} }+\frac{m_{ab}}{2 m_a^2} q^2 - i  \epsilon$ and $f_2(E_2,q,k)=E_2+\frac{q^2}{2(m_a+m_b)}-\frac{k^2}{2 m_b}$.

The agreement between the both sides of (\ref{sigma-app2})  is in general better than $0.1\%$. The approximation in (\ref{sigma-app2}) is used to calculate $\tau_2^{-1}$ and $\Sigma_2$ from (\ref{ts1}) and (\ref{ts}), as shown in figure~\ref{sigmatau}.

%
%

\section{Thresholds of the three-body system} \label{sec:ttbs}

We now treat the problem of two identical impurities of mass $m_a$ immersed in a Fermi sea background and interacting with a fermion of mass $m_b$ that belongs to the sea. 

We assume that the total angular momentum is zero for both interacting and non-interacting impurities. 
As for the two-body system, the interaction between particles is assumed to be attractive $s-$wave zero-range potentials and now the center-of-mass (CM) of the three-body systems is considered at rest in the frame of the Fermi sea. The coupled homogeneous integral equations for the bound state of an $aab$ system are derived in detail in~\cite{bellottiJoPB2011}. Here we have a Fermi sea and need to take Pauli blocking into account. We do so by inserting appropriate $\Theta$-functions in the coupled integral equations.  The Faddeev components describing the system of two impurities immersed in the Fermi sea are given by
\begin{eqnarray}
\fl f_{b}\left( \vec{q}\right) = 2\tau_{aa}\left(E_3-\frac{\vec{q}^{2}}{2 m_{aa,b}}\right) \int
d^2k\frac{\Theta \left( q - k_F \right) f_{a}\left( \vec{k}\right) }{E_{3}-\frac{\vec{q}^{2}}{2 m_{ab}}-\frac{\vec{k}^{2}}{{2 m_{aa}}}-\frac{1}{m_{a}}\vec{k}\cdot
\vec{q}} \ ,  \label{eq.020} \\ 
\fl f_{a}\left( \vec{q}\right)= \tau_{ab}\left(E_3-\frac{\vec{q}^{2}}{ 2m_{ab,a}} \right)\left[ \int d^2k\frac{\Theta \left( k - k_F \right)f_{b}\left( \vec{k}\right)}{E_{3}-\frac{\vec{q}^{2}}{2 m_{aa}}-\frac{\vec{k}^{2}}{2 m_{ab}}-\frac{1}{m_{a}}\vec{k}\cdot \vec{q}} \right. \nonumber\\* \left. 
+\int d^2k\frac{\Theta \left( |\vec{q}+\vec{k}| - k_F \right)
f_{a}\left( \vec{k}\right) }{E_{3}-\frac{1}{2 m_{ab}}\left( \vec{q}^{2}+\vec{k}^{2}\right) -\frac{1}{m_{b}}\vec{k}\cdot \vec{q}}\right] \ ,  \label{eq.021}
\end{eqnarray}
where the three-body reduced masses are $m_{aa,b}=2 m_a m_b/(2 m_a+m_b)$ and $m_{ab,a}= m_a(m_a+ m_b)/(2 m_a+m_b)$. The $ab$ and $aa$ transition amplitudes are respectively given in (\ref{eq06}) and (\ref{eq12}) and are calculated for energies of the corresponding subsystems within the $aab$ system.

Before seeing how the presence of the Fermi sea affects the three-body states, it is important to understand how the two- and three-body thresholds move with the Fermi momentum. As the $a$ particles are not directly affected by the Fermi sea, the transition operator of this subsystem is given in (\ref{eq12}). The two-body breakup due to the interacting identical $a$ particles ($E_{aa}^{th}$), where $aab \rightarrow aa+b$, is defined as the minimum energy which satisfies $\tau_{aa}^{-1}\left(E_{aa}^{th}-\vec{q}^{2}/2 m_{aa,b}\right)=0$. The assumption that the three-body system is at rest in the Fermi sea frame implies that the total momentum of the pair is restricted by the momentum of the fermion ($b$ particle), i.e., $q \geq k_F$. The momentum which minimizes the energy is $k_F$ and $E_{aa}^{th}$ moves with $k_F$ as
\begin{equation}
E^{th}_{aa}=-|E_{aa}| +  \frac{k_F^2}{2 m_{aa,b}}, 
\label{eq13}
\end{equation}
In 2D we set $E_{aa}=0$ when the $a$ particles are not interacting with each other~\cite{bellottiFS2014a,bellottiAe2015}. In this case the two-body transition operator in (\ref{eq.020}) diverges and decouples the homogeneous integral equations ~(\ref{eq.020}) and (\ref{eq.021}). Besides, the $aa+b$ decay channel is not available and $E_{aa}^{th}$ need not be considered in the three-body calculations.

In the same way, the two-body breakup due to the interaction between the $ab$ particles ($E_{ab}^{th}$), where $aab \rightarrow ab+a$, is defined as the minimum energy which satisfies $\tau_{ab}^{-1}\left(E_{ab}^{th}-\vec{q}^{2}/2 m_{ab,a}\right)=0$. As the $b$ particle does interact with the Fermi sea, the transition operator is given in (\ref{eq06}) and the expression which satisfies $\tau_{ab}^{-1}\left(E_2\right)=0$ is presented in (\ref{eq10}). Replacing $E_2 \to E_{ab}^{th}-\frac{\vec{q}^{2}}{2 m_{ab,a}}$ in this equation allows us to calculate the minimum three-body energy that gives a pole in the two-body T-matrix. Notice that there is a subtlety in this part since in cases where $k_F \geq \sqrt{8 m_{ab} |E_{ab}|}$, the minimum energy which defines the two-body breakup does not correspond to an $ab$ bound state (see figure~\ref{bound-virtual}). When the two-body subsystem is virtual (unbound), the three-body system will only be stable while its energy is below the lowest two-body threshold. This condition is found by calculating the expression which minimizes the energy in the expression for $E_{th}^{-}$ when $E_{th}^{-} \to E_{ab}^{th}-\frac{\vec{q}^{2}}{2 m_{ab,a}}$. Collecting everything, the dependence of $E_{ab}^{th}$ with $k_F$ reads
\begin{subnumcases}
{E_{ab}^{th}=\label{eq14} }
-|E_{ab}|+\frac{k_F^2}{2 m_{ab}} 
& $k_F \leq k^* $   \\
-|E_{ab}|- \frac{m_a^2 |E_{ab}|}{m_{ab,a} m_{ab}}+ \sqrt{\frac{2 m_a^2 |E_{ab}|}{m_{ab,a} m_{ab}^2} } k_F 
 & $k^* \leq k_F \leq k^* \left( 1+\frac{ m_{ab} m_{ab,a}}{m_a^2} \right) $   \\
 \frac{k_F^2}{2 m_{ab}} \left(1 - \frac{m_{ab}}{2 m_a}\right) 
& $k_F \geq k^* \left( 1+\frac{ m_{ab} m_{ab,a}}{m_a^2} \right)$ 
\end{subnumcases}
where $k^*= \left(2 m_a^2 |E_{ab}|/m_{ab,a} \right)^{1/2}$.

\begin{figure}[!htb]%
\centering
\includegraphics[width=0.8\columnwidth]{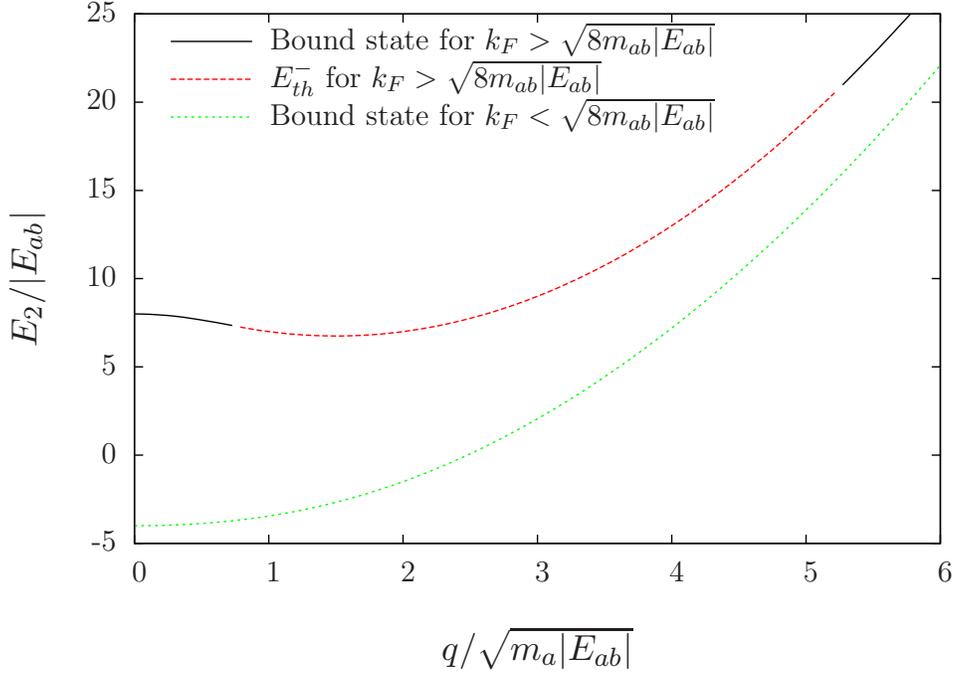}%
\caption{Two-body total energy, $E_2/|E_{ab}|$, as function of $q/\sqrt{m_a |E_{ab}|}$. For $k_F < \sqrt{8 m_{ab} |E_{ab}|}$ the relevant threshold for the three-body $aab$ system is the bound $ab$ subsystem, while for $k_F > \sqrt{8 m_{ab} |E_{ab}|}$ it is the unbound subsystem.}
\label{threshold}
\end{figure}

The result in (\ref{eq14}) is illustrated in figure~\ref{threshold}. The three-body system has to be below either the $ab$ subsystem threshold or the two-body continuum when the first one is not available.  

Finally, the three-body breakup ($E_{3B}^{th}$), where $aab \rightarrow a+a+b$, is defined as the minimum energy which renders one of the denominators on the right-hand-side of (\ref{eq.020}) and (\ref{eq.021}) zero. For $k_F=0$ this energy is $E_{3B}^{th}=0$ and a finite $k_F$ increases the breakup value. Searching for the minimum energy which makes the denominator vanish on the first term of both (\ref{eq.020}) and (\ref{eq.021}), we have that one of the momenta $\vec{q}$ or $\vec{k}$ is allowed to vary in the range $[0,\infty]$, while the $\Theta-$function restrict the other one to $[k_F,\infty]$. In both cases, the minimum energy that makes the denominator zero is $E_{3B}^{th}=k_F^2/2 m_{ab}$. In the missing term, the second one on the right-hand-side of (\ref{eq.021}), both momenta are allowed in the range $[0,\infty]$, but the $\Theta-$function restricts this whole term to contribute only when $q+k \geq k_F$ (see (\ref{theta2a})). The dependence of $E_{3B}^{th}$ on $k_F$ is therefore
\begin{equation}
E_{3B}^{th} = \frac{k_F^{2}}{4 m_{a}}  , 
\label{eth}
\end{equation}  
which is always below $k_F^2/2 m_{ab}$. \Eref{eth} thus defined the three-body threshold for the breakup $aab\rightarrow a+a+b$.

\section{Interacting impurities} \label{sec:ii}

\begin{figure}[!htb]%
\centering
\includegraphics[width=0.8\columnwidth]{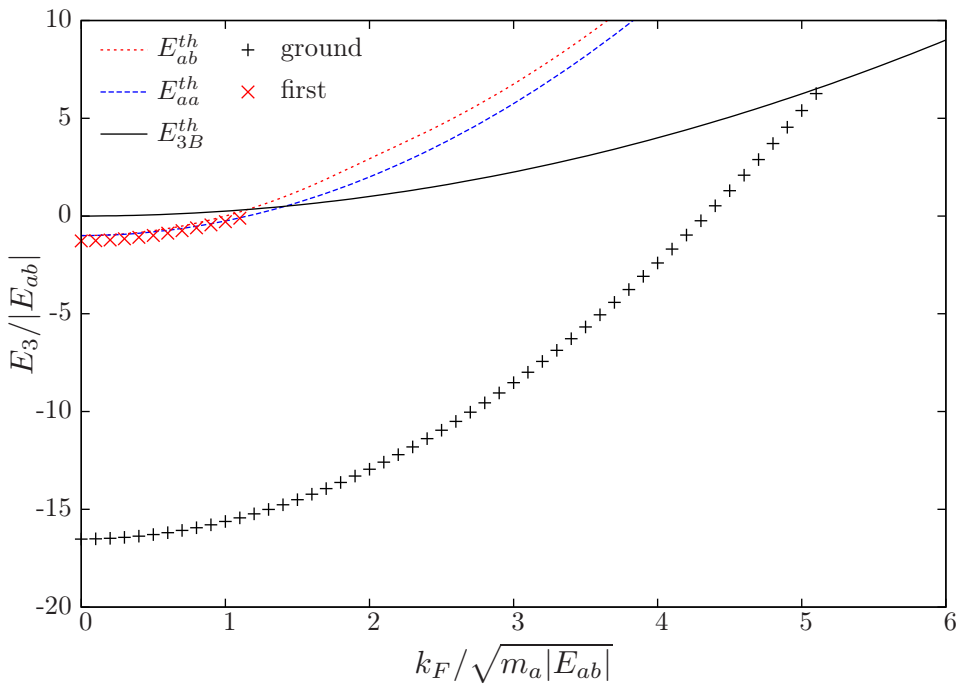}
\caption{Three-body energy, $E_3/|E_{ab}|$, as function of the Fermi momentum, $k_F/\sqrt{m_a |E_{ab}|}$, for $m_a=m_b$ and $E_{aa}=E_{ab}$ The $\downarrow$ marks the point $k_F = k^*$. For $k_F > k^*$ the two $ab$-subsystems are unbound.}
\label{mequ1}
\end{figure}

Each impurity particle, $a$, interacts with a fermion from the sea by a zero-range potential characterized by a two-body binding energy $E_{ab}=\left(2 m_{ab} a_{2D}^{2}\right)^{-1}$, forming three-body bound states $aab$. These states can exist even when the impurities are not interacting with each other. In both the interacting ($E_{aa}\neq0$) and non-interacting ($E_{aa}=0$) cases, the three-body energies increase with increasing Fermi momentum $k_F$. Since the two- and three-body thresholds are also moving up as $k_F$ increases, three-body bound states are found below the Fermi energy even with positive energies. As an example, we show in figure~\ref{mequ1} the case where the two impurities have the same mass as the fermion ($m_a=m_b$). 

The two impurities can, in principle, interact with each other with any energy. We set $E_{aa}=E_{ab}$, which for identical masses and $k_F=0$ leads to the well-known case of just one two-body state and two three-body bound states with energies $E_3^{(0)}=16.52E_2$ and $E_3^{(1)}=1.27E_2$~\cite{bruchPRA1979,bellottiJoPB2011}. As seen in figure~\ref{mequ1}, the energies and thresholds increase with increasing $k_F$. For $E_{aa}=E_{ab}$, the decay channel $aab \to a+ab$ is always closed. The three-body threshold is highest for $k_F=0$ and moves up at the smallest rate. It goes below $E_{aa}^{th}$ when $k_F/\sqrt{m_a |E_{ab}|}>\sqrt{2}$. It is interesting to note that the first excited state decays into an atom and a dimer when $k_F/\sqrt{m_a |E_{ab}|}\approx 1.1$, while the ground state is breaking up into three free atoms for $k_F/\sqrt{m_a |E_{ab}|}\approx5.1$.

In view of what was discussed in the previous section, for $k_F \geq k^* = \left(2 m_a^2 |E_{ab}|/m_{ab,a} \right)^{1/2}$ three-body states of the two interacting impurities, which in 2D forms a L=0 bound state,  and a fermion from the sea are supported even when two of the subsystems are in the virtual state, as shown in figure~\ref{mequ1}.

The results in figure~\ref{mequ1} indicate that cold atomic experiments may be able to deepen the knowledge of 2D three-body systems by looking at two impurities in the presence of a Fermi sea with which the impurities interact. The Fermi momentum is given by density of the sea, which is a controllable parameter. Then, it is possible to change it and measure the number of atoms lost in the trap, which must have a peak when the three-body system is close to either the two- or three-body continuum~\cite{kraemerNP2006}.  Different bound states may decay into different systems, as seen in figure~\ref{mequ1} and a larger number of bound states can be reached in highly asymmetric mass systems ($m_b/m_a\ll1$)~\cite{bellottiJoPB2013}. 

\begin{figure}[!htb]%
\centering
\includegraphics[width=0.8\columnwidth]{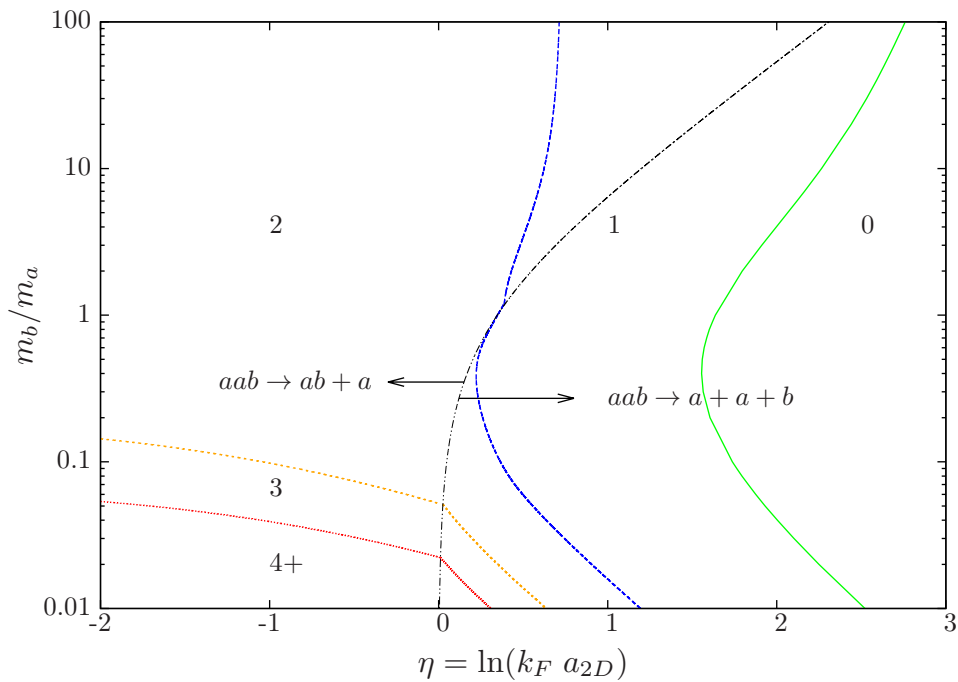}
\caption{Bound state diagram for $E_{aa}=E_{ab}$ as function of the interaction strength $\eta$. The Arabic numerals indicate the number of bound states in each region. The ``+'' sign in $4+$ indicates that more than four bound states can be found in that region. The central black-line divides the plot in two main regions where the decay channel is either $aab \to b+aa$ or $aab \to a+a+b$.}%
\label{diagrams}%
\end{figure}

In order to give some insight for experiments, it is possible to systematically extend the procedure used in the investigation of the symmetric mass case and study how the three-body states vanish for a large range of mass-ratios.
The final result is summarized in a mass versus $k_F$ diagram, shown in figure~\ref{diagrams} for $E_{aa}=E_{ab}$. The Arabic numerals indicate the number of bound states in each region, where 0 means that no bound state was found, 1 indicates that only the ground state is available and so on. The central black-line divides the plot in two main regions where the three-body states go into either a dimer plus atom or three-atom continuum. Taking as example two $^{133}$Cs atoms immersed in a sea of $^{6}$Li fermions, the mass-ratio is $m_b/m_a=0.045$ and four bound states are present for both $E_{aa}=0$ and $E_{aa}=E_{ab}$ when $k_F=0$~\cite{bellottiFS2014,bellottiPRA2012}. It is shown in figure~\ref{diagrams} that the four states for $E_{aa}=E_{ab}$ disappears at $\eta\approx 1.96$ (ground), $\eta\approx 0.59$ (first), $\eta\approx 0.058$ (second) and $\eta\approx -1.42$ (third), respectively. 
Furthermore, notice that the three deeper states decay into three atoms while the highest one decays into atom plus dimer.

The result shown in figure~\ref{diagrams} is, to the best of our knowledge, the first one to consider the problem of two impurities immersed in a Fermi sea in 2D when the impurities are allowed to interact with each other. These results were achieved without taking into account particle-hole fluctuations in the Fermi sea. We now consider fluctuation corrections starting with two non-interacting impurities.

\section{Self-energy corrections to the three-body system} \label{sec:niisectbs}

\begin{figure}[!htb]%
\centering
\includegraphics[width=0.8\columnwidth]{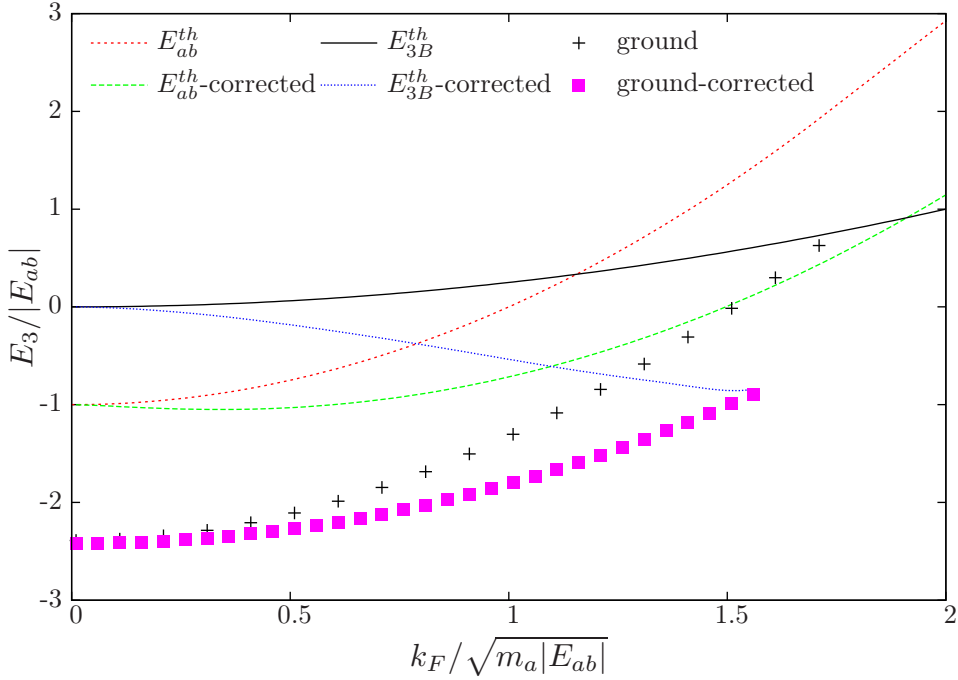}
\caption{Three-body energy, $E_3/|E_{ab}|$, as function of the Fermi momentum, $k_F/\sqrt{m_a |E_{ab}|}$, for $m_a=m_b$ and $E_{aa}=0$. Results with and without the self-energy correction are compared. }
\label{mequ0}%
\end{figure}

Here we study the three-body system for two non-interacting impurities and then we see how the fluctuations in the Fermi sea affect the results. Taking $E_{aa}=0$ leads to $f_b(\vec{q})=0$ in (\ref{eq.020}) and (\ref{eq.021}), simplifying the calculation. Furthermore, for $m_a=m_b$ and $k_F=0$ only one bound state is supported~\cite{bellottiFS2014} as shown in figure~\ref{mequ0}. Without considering the particle-hole fluctuations in the Fermi sea, the behavior of the three-body energy and of the thresholds is similar to the case where the impurities are allowed to interact. For $k_F/\sqrt{m_a |E_{ab}|}>0$, the three-body threshold, which is the highest for $k_F=0$, increases at the smallest rate and for $k_F/\sqrt{m_a |E_{ab}|}>1.155$ it becomes lower than the $ab$ threshold, meaning that the decay channel $aab \to a+ab$ is not available any more for the three-body system. When $k_F/\sqrt{m_a |E_{ab}|}\approx1.7$ the three-body state touches the continuum and the system decays into three free atoms. 

As in the case of interacting impurities, for $k_F \geq k^* = \left(2 m_a^2 |E_{ab}|/m_{ab,a} \right)^{1/2}$ the three-body ground state is supported even when the two $ab$-subsystems are in a virtual state. Since the last pair is also unbound (non-interacting impurities), the three-body system is bound when the three subsystems are unbound. The behavior of the ground state is similar to what has been seen in so-called Borromean systems in 2D~\cite{volosnievEPJD2013,volosnievJoPBAMaOP2014}. In the present case we have the background medium in the form of a Fermi sea which is what makes this behavior possible. One may therefore consider this an example of a medium-induced Borromean state in 2D. 

Next we introduce the self-energy correction in the three-body equations. Since the two impurities are not interacting, only the second term on the right-hand-side of (\ref{eq.021}) survives. It is important to emphasize at this point that the bosons in our problem are not from a Bose gas, but isolated impurities, or we can say that the average distance with a third boson from the gas is very large compared to any size scales in the problem (vanishing limit of the density). In this idealized situation, the connected three-body T-matrix can in principle contribute only to the two-boson T-matrix with a term proportional to the fermion density. If in addition, we take into account that the Bose gas density is negligible according to our statement of the physical situation, the contribution of the connected three-body T-matrix to the self-energy will be negligible and only the non-connected term of the T-matrix with the fermion and boson T-matrix will be relevant. Indeed, this term was already taken into account in our evaluation of the self-energy of the impurity.  In a more general situation, e.g., in the presence of a Bose gas, other contributions to the impurity self-energy should be included, and the feedback from the full three-body T-matrix should be relevant as the boson density increases. In this case, one has to include corrections to the self-energy beyond the contribution of the disconnected part of the three-body transition matrix.

That said, the first correction comes in the two-body sector where the transition operator in (\ref{eq06}) is replaced by  the one in (\ref{ts}) with $n=0$. As we discussed before, the convergence of the two-body T-matrix with $n$ is very fast and the most significant correction is seen in figure~\ref{sigmatau} to be from $\tau_0^{-1}$ to $\tau_1^{-1}$, justifying the choice of $n=0$ in (\ref{ts}). In order words, the real gain of going to higher $n$ is not worth the extra time spent in the calculation. Then, considering the dressed propagator of the impurity, the integral equation is written as
\begin{eqnarray}
\fl f(\vec{q})=-\tau_1\left(E_3-\frac{\vec{q}^{2}}{ 2m_{ab,a}} \right) \int \frac{d E_2}{2 \pi i} 
\frac{d^2 k}{E_2-\frac{k^2}{2 m_a} -\Sigma_a(E_2,\vec{k})+ i  \epsilon}  \nonumber\\*
\times \frac{ \Theta\left( \left| \vec{q}+\vec{k} \right| -k_F\right) f(\vec{k}) }{E_3-E_2-E_a-\frac{(\vec{q}+\vec{k})^2}{2 m_b} + i  \epsilon}  \; ,
\label{fsig}
\end{eqnarray}
with $\tau_1^{-1}$ given in (\ref{ts}). The numerical solution of (\ref{fsig}) gives the corrected three-body energy and we need to understand again how the thresholds and energies depends on the Fermi momentum. However, since the self-energy correction is being considered, both the two- and three-body corrected thresholds are calculated numerically.

As shown in figure~\ref{sigmatau}, the inclusion of the self-energy in the propagator of the impurity renders the molecule more bound, which lowers the $ab$ threshold in the three-body calculation (see figure~\ref{mequ0}). The same happens to the three-body threshold, which starts from zero when $k_F=0$ and goes negative for $k_F>0$. The three-body state is also more bound and 
its energy as function of $k_F$ increases slower than in the previous cases, where corrections due to fluctuations of the medium were not considered. 
The three-body energy and the thresholds with and without the self-energy correction are compared in figure~\ref{mequ0}. The final result is that, although rendering both the two- and three-body bound states more bound, the inclusion of the self-energy makes the states disappear at a slightly smaller $k_F$, since the changing rates of the thresholds are strongly affected.

\begin{figure}[!htb]%
\centering
\includegraphics[width=0.8\columnwidth]{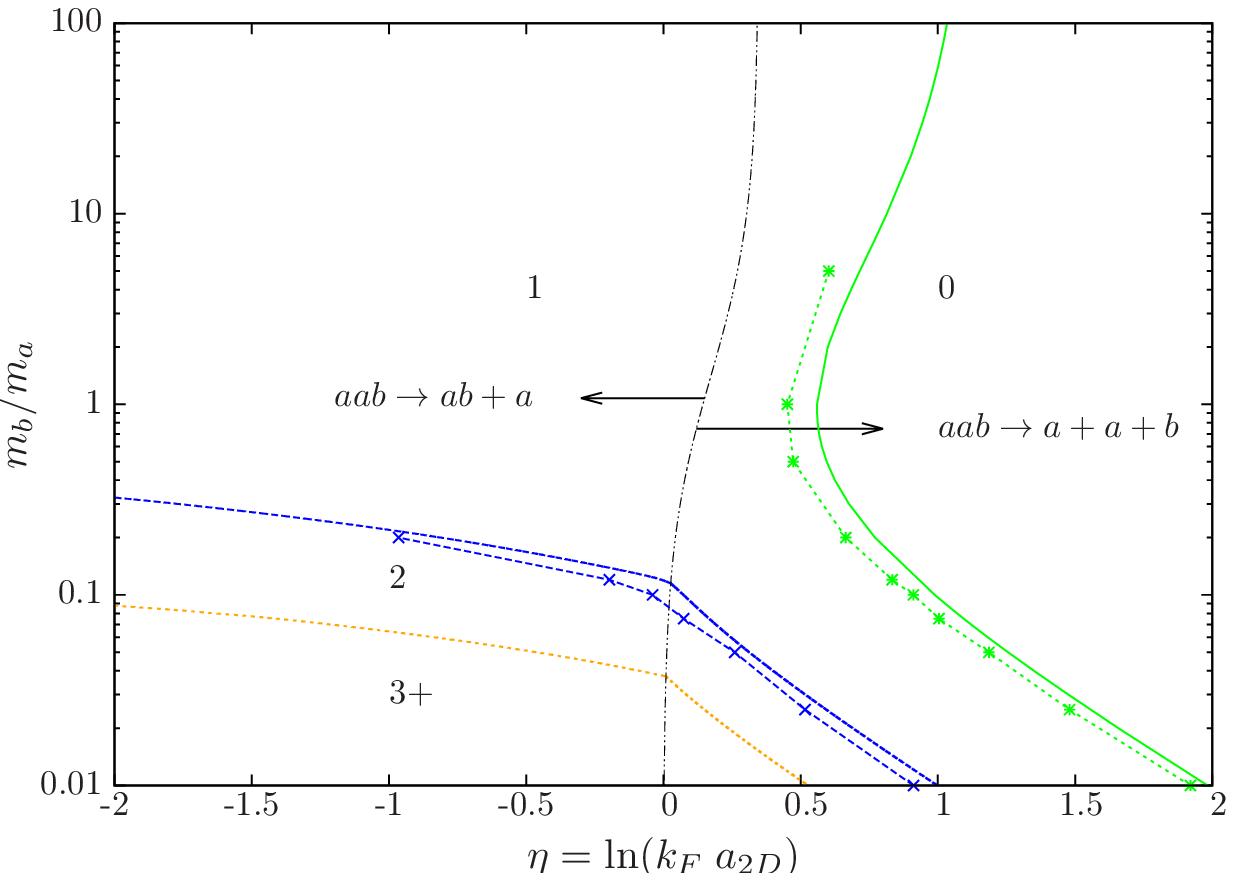}
\caption{Bound state diagram for $E_{aa}=0$ as function of the interaction strength $\eta$. The Arabic numerals indicate the number of bound states in each region. The ``+'' sign in $3+$ indicates that more than three bounds states can be found in that region. The central black-line divides the plot in two main regions where the decay channel is either $aab \to a+ab$ or $aab \to a+a+b$.}%
\label{diagram_corrected}%
\end{figure}

The extension of the result in figure~\ref{mequ0} for another mass ratios gives the diagram shown in figure~\ref{diagram_corrected}, where the Arabic numeral indicate the number of bound states in each region. The continuous lines are from calculations without fluctuations in the Fermi sea and the discrete points show how the lines move when the self-energy is considered. Notice that the correction is always small within the range of the Fermi momentum and masses considered, which covers typically experimentally accessible cold atomic gas systems. 
Such correction is expected to be smaller in the case of interacting impurities, where the extra attraction would render the effects of the Fermi sea less relevant.

\section{Discussion and outlook} \label{sec:do}
We have considered the 2D problem of two identical atomic impurities (either bosonic or with distinct internal states), immersed in a background Fermi sea and interacting with a fermion on top of it in 2D. The interactions were modeled by attractive zero-range potentials and the Faddeev decomposition was used to write the homogeneous coupled integral equations for the three-body bound state.

The problem of just one impurity propagating in the Fermi sea was first considered and we have reviewed some well-known results in the literature~\cite{zollnerPRA2011, parishPRA2011, schmidtPRA2012,parishPRA2013,ngampruetikornEEL2012}, specifically the two-body T-matrix in the medium (\ref{eq06}) and the energy of the molecule as function of its total momentum (\ref{eq10}). The binding energy of the molecular state was studied previously as a function of its mass ratio, total momentum, and Fermi energy. For some values of these parameters the pole of the T-matrix, which represents a bound state, enters through the cut of the two-body continuum and increasing the momentum of the molecule it returns from the cut and become a bound state again~\cite{schmidtPRA2012,EngelbrechtPRB1992}. We have shown here that this state appears in the second energy sheet and becomes a virtual state, as shown in figure~\ref{bound-virtual}. 
Replacing $E_2 \to E_2+ i \epsilon$ in order to consider the two-body T-matrix in the medium  (\ref{eq06}) in the scattering region leads to inconsistent results in some cases (see figure~\ref{tau}). We have shown that the analytic extension of these matrix elements, as presented in (\ref{eq08}), gives the right solution in this case.

Particle-hole fluctuations were then considered and the self-energy of the impurity propagating in the sea was self-consistently calculated. The non-linear couped equations for the transition operator and the self-energy (\ref{ts1}) and (\ref{ts}) were solved iteratively and we showed that the convergence of both quantities with the number of iterations was very fast. Although this method has some disadvantages, we found it suitable to use mainly in the three-body calculation. 
This choice is supported since the self-consistent method employed in this work correctly describes the two-body system when particle-hole fluctuations are taken into account. We found that the energy of the attractive and repulsive branches of the polaron, as well as their weights and the polaronic spectral function are the same as found in~\cite{ngampruetikornEEL2012,schmidtPRA2012,LevinsenPRA2012}. The polaron energy is the relevant information for $\eta>-0.54$, where most of the effects of the medium on the three-body system are calculated (see figures \ref{diagrams} and \ref{diagram_corrected}).

The formalism used here allowed us to study the complex problem of two interacting impurities propagating in the Fermi sea. 
Interestingly, three-body states of the two interacting impurities and a fermion of the sea are supported even when two of the subsystems are in a virtual state. Importantly, the fate of the three-body state depends strongly on the two-body systems. 
This feature can be used to identify the three-body states of two impurities immersed in a Fermi sea in measurements of cold atoms lost from a trap.

The complexity of the integral equations when the impurities are allowed to interact makes the inclusion of the self-energy correction very hard to be implemented. However, if the impurities are not interacting, the integral equations simplify and the effect of the correction can be studied. First of all, the three-body bound states are still allowed when the impurity-fermion subsystems are in a virtual state. Since we have two non-interacting impurities, the three-body systems are bound when the three two-body subsystems are unbound,
which can be interpreted as a medium-induced Borromean state in 2D. The fate of the three-body states once more strongly depends on the two-body subsystems. The inclusion of the self-energy of the impurity drastically affects the behavior of the subsystems and of the three-body states, as it can be seen in figure~\ref{mequ0}, but the superposition of all the effects together leads to just a small correction in the final results, as shown in figure~\ref{diagram_corrected}. We expect that the correction would be smaller in the case of interacting impurities, since the effects of the Fermi sea would decrease as there is more attraction in the system.

Three-body bound states were also found when a single impurity is immersed in a Fermi sea~\cite{parishPRA2013}. The different particle can be in a polaronic state or bind one or two fermions from the sea. It is interesting to understand what scenario is more favorable when another impurity is brought into the game. Would each impurity bind to two fermions or the two impurities bind to one fermion? Valuable information can also be gained by considering two polarons interacting with each other as a four-body system of two fermions and two impurities, using techniques similar to the ones employed in~\cite{hadizadehPRL2011}. Another interesting direction for future investigations would be to study how the presence of the Fermi sea would affect the momentum distribution of the three-body systems by calculating the measurable two- and three-body contacts through an extension of the techniques described in~\cite{bellottiNJoP2014}. Another interesting point is the understanding of how the signature of the Efimov effect in 3D three-body systems under the influence of a Fermi sea~\cite{nygaardNJoP2014} would disappear and connect to the result presented here through a change in dimensionality, similarly to what was done for three-identical bosons in~\cite{levinsenPRX2014,yamashitaJoPB2015}.

\vskip 1em

\paragraph*{Acknowledgements}
The authors thank partial support from the Brazilian agencies
FAPESP, CNPq and CAPES (88881.030363/2013-01), and by the Danish
Council for Independent Research DFF Natural Sciences
and the DFF Sapere Aude program.

\appendix  
\section{Equation for the self-energy} \label{sec:ese}
The bare and full propagators and the self-energy of the impurity of mass $m_a$ immersed in a sea of fermions $m_b$ relates to each other through the Dyson equation $G_\Sigma^{-1}(E,\vec{p})=E-\frac{p^2}{2m_a}-\Sigma_a(E,\vec{p})$.
In the self-consistent way, the transition operator is written as
\begin{eqnarray}
\fl \tilde{\tau}(E_a,p_a,E_a',p_a';E_b,p_b,E_b',p_b')= i \lambda+ \left(i  \lambda\right)^2 I(E,q)+\left(i  \lambda\right)^3 I^2(E,q)+ ... \nonumber\\*
= \frac{i }{\lambda^{-1}-I(E,q)}  \; , 
\label{tautilde} 
\end{eqnarray} 
where the total energy and total momentum are respectively $E=E_a+E_b$ and $\vec{q}=\vec{p}_a+\vec{p}_b$. The integral $I(E,q)$ reads
\begin{equation}
\fl I(E,q)=\int{\frac{d E'_b}{2 \pi} \; d^2 k 
\frac{\Theta\left( k-k_F\right)}{E-E'_b-\frac{(\vec{q}-\vec{k})^2}{2 m_a}-\Sigma_a(E-E'_b,\vec{q}-\vec{k}) + i  \epsilon} }
\frac{1}{E'_b-\frac{k^2}{2 m_b} + i  \epsilon} 
\label{ieq}
\end{equation}
and the expression for the self-energy is found to be
\begin{equation}
\Sigma_a(E_a,\vec{p}_a)
 = - \int{d^2 p_b \; \frac{\Theta\left( k_F -p_b\right)}{\lambda^{-1}-I(E_a+\frac{\vec{p}_b^2}{2 m_b},\vec{p}_a+\vec{p}_b)} } \; . 
\label{sigmaa} 
\end{equation} 

The expressions for the transition amplitude and the self-energy, given respectively in (\ref{tautilde}) and (\ref{sigmaa}), form a system of coupled equations. We solve this system iteratively by firstly setting the zeroth-order of the self-energy in (\ref{sigmaa}) as
\begin{equation}
\Sigma_a^0(E_a,\vec{p}_a)=\int_{p_b<k_F}{d^2p_b \; \tau_0\left(\left|\vec{p}_b+\vec{p}_a\right|,\frac{p_b^2}{2 m_b}+E_a\right)} \;,
\label{self-energy}
\end{equation} 
where the zeroth-order transition operator $\tau_0$ is the one given in (\ref{eq08}) and is independent of $\Sigma_a$. Then, $\Sigma_a^0$ is introduced back in (\ref{tautilde}), leading to the first-order operator $\tau_{1}$, which requires (\ref{ieq}) to be solved. 
Performing the integral on the energy in (\ref{ieq}) is not trivial and an analysis of the poles has to be done. In the complex plane of $E'_b$, the poles of $I(E,q)$ are
\begin{eqnarray}
E'_b=\frac{k^2}{2 m_b} - i  \epsilon \ , \label{ebpole1} \\
E'_b=E-\frac{(\vec{q}-\vec{k})^2}{2 m_a}-\Sigma_a(\vec{q}-\vec{k},E-E'_b) + i  \epsilon \ .  \label{ebpole2}
\end{eqnarray} 
The pole in (\ref{ebpole1}) is always on the lower half complex plane of $E_b'$, as the only imaginary part comes from the small contribution $i  \epsilon$ and, for $\Sigma_a=0$, the pole in (\ref{ebpole2}) is on the upper half plane. The self-energy is complex and contributes to the location of the pole. The study of its behavior and the results for the polaron spectral function in~\cite{schmidtPRA2012} show that $\Im(\Sigma_a) \leq 0$ for any $(k,q,E,E'_b)$, ensuring that the pole in (\ref{ebpole2}) is on the upper-half complex plane of $E_b'$ even for $\Sigma_a \neq 0$.

It is also necessary to study the analyticity of $\tau$ and $\Sigma_a$ as function of $E$ and $E_b'$. The coupled equations (\ref{ieq}) and (\ref{sigmaa}) indicates that both $\tau$ and $\Sigma_a$ must be analytic in the same region. From (\ref{eq01}) is possible to see that $\tau$, and consequently $\Sigma$, are analytic on the upper half complex plane of $E$ and therefore on the lower half complex plane of $E_b$. 
As the pole in (\ref{ebpole1}) is in the semi-plane where the functions $\tau$ and $\Sigma_a$ are analytic, the contour is closed through the lower half complex plane of $E'_b$ (counterclockwise). The integral in (\ref{ieq}) is calculated using the residues theorem  and leads to the transition operator (see (\ref{tautilde})) 
\begin{eqnarray}
\fl \tau^{-1}_{1}(E_2,q)=\lambda^{-1} \nonumber\\* \hskip -5em
-\int{  
\frac{d^2k \ \Theta \left( k - k_F \right)}{-E_2 +\frac{k^2}{2 m_{ab} }+\frac{m_{ab}}{2 m_a^2} q^2+ \frac{k q}{ m_a} \cos\theta + \Sigma_a^0(\vec{q}-\vec{k},E_2+\frac{q^2}{2(m_a+m_b)}-\frac{k^2}{2 m_b}) - i  \epsilon}} \;  ,
 \label{tauc} 
 \end{eqnarray}
where $\lambda^{-1}$ is related to the energy of the pair in the vacuum (see (\ref{eq01})). The set of decoupled equations for the self-energy and the transition operator are
\begin{eqnarray}
\fl \Sigma_a^n(E_a,\vec{p}_a)=\int_{p_b<k_F}{d^2p_b \; \tau_n\left(\left|\vec{p}_b+\vec{p}_a\right|,\frac{p_b^2}{2 m_b}+E_a\right)} \; , \; n \geq 0  \label{ts1-app}\\
\fl \tau^{-1}_{n+1}(E_2,q)=\lambda^{-1} \nonumber\\* \hskip -5em
-\int{ d^2k 
\frac{\Theta \left( k - k_F \right)}{-E_2 +\frac{k^2}{2 m_{ab} }+\frac{m_{ab}}{2 m_a^2} q^2+ \frac{k q}{ m_a} \cos\theta + \Sigma_a^n(\vec{q}-\vec{k},E_2+\frac{q^2}{2(m_a+m_b)}-\frac{k^2}{2 m_b}) - i  \epsilon}} \ ,  \label{ts-app}
\end{eqnarray} 
where $\tau_0$ is given in (\ref{eq08}) and is independent of $\Sigma_a$.

\section{Angular part of the integral Equation} \label{sec:apie}
Since we are interested in states with total angular momentum zero the spectator functions in (\ref{eq.020}) and (\ref{eq.021}) are independent of the angle between the momenta $\vec{q}$ and $\vec{k}$. The angular dependence of the first term on the right-hand-side of these equations comes only from the denominator, as both the spectator ($f(\vec{q})\equiv f(q)$) and the $\Theta-$functions are angle-independent.  

However, the $\Theta-$function in the second term of (\ref{eq.021}) does depend on the angle, which makes the integration of such term not as simple as the previous two. The trick to solve this integral is to transfer the information of the $\Theta-$function to the limits of the angular integral. The angular dependence in the $\Theta-$function on the second term of (\ref{eq.021}) is
\begin{equation}
k^2+q^2+2kq\cos\theta  \geq k_F^2 \; , 
\label{ang3}
\end{equation}
which defines the angle $\theta_1$ as the lowest limit of $\theta$ in (\ref{ang3}) as 
\begin{equation}
\cos\theta_1=\frac{k_F^2-k^2-q^2}{2kq} \;  
\end{equation}
with $0\leq\theta_1\leq\pi$.

Changing integration limit from $2 \pi$ to $\theta_1$, the angular integral in the second term of (\ref{eq.021}) reads     
\begin{equation}
2 \; \int_0^{\theta_1}{ \frac{d\theta}{1+a \cos\theta}}
=\frac{ 4 \tan^{-1} \left[ \sqrt{\frac{1-a}{1+a}} \tan \left(\frac{\theta_1}{2}\right) \right]}{\sqrt{1-a^2}} \; ,
\label{ang2}
\end{equation}
where the main branch of $\tan \theta$ has to be considered. Notice that for $\theta_1=\pi$, (\ref{ang2}) gives the result for the angular integration of the first term on the right-hand-side of (\ref{eq.020}) and (\ref{eq.021}).

The dependence of the $\Theta-$ function on the angle is eliminated using that $|\cos \theta_1| \leq 1$. The result is 
\begin{subnumcases}
{\int_0^{2 \pi}{ \frac{d\theta \; \Theta \left( |\vec{q}+\vec{k}| - k_F \right)}{1+a \cos\theta}}=} 
0  & $q+k \leq k_F$    \label{theta1a} \\
\nonumber \\
\frac{2 \pi}{\sqrt{1-a^2}}  & $|q-k| \geq k_F$  \;  . \label{theta1b} \\
\nonumber \\
\frac{ 4 \tan^{-1} \left[ \sqrt{\frac{1-a}{1+a}} \tan \left(\frac{\theta_1}{2}\right) \right]}{\sqrt{1-a^2}}  & elsewhere \label{theta1c}
\end{subnumcases}

After the angular integration, the homogeneous coupled integral (\ref{eq.020}) and (\ref{eq.021}) read
\begin{eqnarray}
\fl f_{b}\left( q \right) = -4 \pi \tau_{aa}\left(E_3-\frac{q^{2}}{2 m_{aa,b}}\right)  \int_0^\infty
 \frac{ dk \; k \; \Theta \left( q - k_F \right) f_{a}\left( k \right) }{\sqrt{\left(-E_{3}+\frac{q^{2}}{2 m_{ab}}+\frac{k^{2}}{2 m_{aa}}\right)^2-\frac{k^2 q^2}{m_{a}^2}}}, & \label{eq.020a} \\ 
\fl f_{a}\left( q\right)= -2 \pi \tau_{ab}\left(E_3-\frac{q^{2}}{2m_{ab,a}}\right) \left[  \int_{k_F}^\infty \frac{dk \; k \; f_{b} \left(k \right)}{\sqrt{\left(-E_{3}+\frac{q^{2}}{2 m_{aa}}+\frac{k^{2}}{2 m_{ab}}\right)^2-\frac{k^2 q^2}{m_{a}^2}}} \right.  \nonumber\\
 \left. +\int_0^\infty \frac{dk \; k \; g \left( q,k,k_F \right) \; f_{a}\left(k \right)}
{\sqrt{\left(-E_{3}+\frac{1}{2 m_{ab}}\left( q^{2}+k^{2}\right)\right)^2 -\frac{k^2q^2}{m_{b}^2} }}\right] \ ,   \label{eq.021a}
\end{eqnarray}
where the function $g(q,k,k_k)$ reads
\begin{subnumcases}
{} 
0  & $q+k \leq k_F$    \label{theta2a} \\
\nonumber \\
1  & $|q-k| \geq k_F$  \;  . \label{theta2b} \\
\nonumber \\
\frac{ 2 }{\pi} \tan^{-1} \left[ \sqrt{\frac{-E_{3}+\frac{1}{2 m_{ab}}\left( q^{2}+k^{2}\right) -\frac{1}{m_{b}} kq}{-E_{3}+\frac{1}{2 m_{ab}}\left( q^{2}+k^{2}\right) + \frac{1}{m_{b}} kq}} \tan \left(\frac{\theta_1}{2}\right) \right]  & elsewhere \label{theta2c}
\end{subnumcases}

\end{document}